\documentclass[runningheads,bookmarksdepth=2]{llncs}

\usepackage[T1]{fontenc}

\usepackage{graphicx}

\usepackage[svgnames,dvipsnames]{xcolor}

\usepackage{enumitem}

\usepackage{xspace}

\usepackage{tikz}
\usetikzlibrary{arrows.meta}
\usetikzlibrary{shapes.geometric}

\usepackage{hyperref}

\usepackage[nameinlink,noabbrev]{cleveref}
\newcommand{\etal}{\textit{et al.}}

\newcommand{\OG}{\textsc{Ocean Guard}\xspace}

\newcommand{\Requirements}[1]{\textit{#1}\xspace}
\newcommand{\Actor}[1]{\textsc{#1}\xspace}
\newcommand{\Team}[1]{\textit{#1}\xspace}
\newcommand{\Component}[1]{\texttt{#1}\xspace}

\definecolor{drawio-blue}{HTML}{6c8ebf} %
\definecolor{drawio-gray}{HTML}{666666} %
\definecolor{drawio-green}{HTML}{82b366} %
\definecolor{drawio-orange}{HTML}{fa6800} %
\definecolor{drawio-pink}{HTML}{99004D} %
\definecolor{drawio-purple}{HTML}{9673a6} %
\definecolor{drawio-red}{HTML}{b85450} %
\definecolor{drawio-white}{HTML}{f9f7ed} % 
\definecolor{drawio-black}{HTML}{36393d} % 
\definecolor{drawio-yellow}{HTML}{d6b656} %

\definecolor{drawio-violet}{HTML}{6a00ff} %
\definecolor{drawio-magenta}{HTML}{dd0073} %
\definecolor{drawio-moss}{HTML}{008a00} %

\newcommand{\TikzSkewedSquare}[1][black]{%
  \tikz[baseline] {\draw[transform shape, black, fill={#1}] (0,0) rectangle (1.5ex,1.5ex);}%
}
\newcommand{\TikzSkewedCircle}[1][black]{%
  \tikz[baseline] {\draw[yshift=0.75ex, anchor=south, transform shape, black, fill={#1}] (0,0) circle (0.75ex);}%
}
\newcommand{\TikzSkewedHexagon}[1][black]{%
  \tikz[baseline] {\node[yshift=0.75ex, regular polygon, regular polygon sides=6, shape border rotate=0, draw=black, fill={#1}, scale=0.8] (hex) {}}%
}

\newcommand{\LegendColoredSquare}[3]{%
  \textcolor{#1}{#3}~\TikzSkewedSquare[#2]%
}
\newcommand{\LegendColoredCircle}[3]{%
  \textcolor{#1}{#3}~\TikzSkewedCircle[#2]%
}
\newcommand{\LegendColoredHexagon}[3]{%
  \textcolor{#1}{#3}~\TikzSkewedHexagon[#2]%
}

\newcommand{\LegendColoredComponent}[2]{%
  \LegendColoredSquare{drawio-#1}{drawio-#1}{#2}%
}
\newcommand{\LegendBWComponent}[1]{%
  \LegendColoredSquare{drawio-black}{drawio-white}{#1}%
}

\newcommand{\LegendColoredLabel}[2]{%
  \LegendColoredCircle{drawio-#1}{drawio-#1}{#2}%
}
\newcommand{\LegendColoredLayer}[2]{%
  \LegendColoredHexagon{drawio-#1}{drawio-#1}{#2}%
}

\newcommand{\LegendService}[2]{%
  \textcolor{drawio-violet}{{#2~(#1)}}%
}
\newcommand{\LegendPipeline}[2]{%
  \textcolor{drawio-moss}{{#2~(#1)}}%
}
\newcommand{\LegendDataStore}[2]{%
  \textcolor{drawio-magenta}{{#2~(#1)}}%
}

\hypersetup{
  bookmarksopen=true,    % Show bookmarks bar?
  pdftitle={MLOps with Microservices: A Case Study on the Maritime Domain},
  pdfauthor={R. C. Ferreira et al.},
  pdfcreator={pdflatex}, % Creator of the document
  pdfnewwindow=true,     % Links in new window
  colorlinks=true,       % false: boxed links; true: colored links
  linkcolor=blue,        % Color of internal links
  citecolor=DarkGreen,   % Color of links to bibliography
  filecolor=green,       % Color of file links
  urlcolor=DarkRed       % Color of external links
}

\begin{document}

%%%%%%%%%%%%%%%%%%%%%%%%%%%%%%%%%%%%%%%%%%%%%%%%%%%%%%%%%%%%%%%%%%%%%%%%%%%%%%%%
%                                    TITLE                                     %
%%%%%%%%%%%%%%%%%%%%%%%%%%%%%%%%%%%%%%%%%%%%%%%%%%%%%%%%%%%%%%%%%%%%%%%%%%%%%%%%

% Title in the first page (breaking line after :)
\title{MLOps with Microservices:\break A Case Study on the Maritime Domain}
% Title in the header (no line breaks)
\titlerunning{MLOps with Microservices: A Case Study on the Maritime Domain}

%%%%%%%%%%%%%%%%%%%%%%%%%%%%%%%%%%%%%%%%%%%%%%%%%%%%%%%%%%%%%%%%%%%%%%%%%%%%%%%%
%                                   AUTHORS                                    %
%%%%%%%%%%%%%%%%%%%%%%%%%%%%%%%%%%%%%%%%%%%%%%%%%%%%%%%%%%%%%%%%%%%%%%%%%%%%%%%%

\author{
Renato Cordeiro Ferreira\inst{1,2,3}\orcidID{0000-0001-7296-7091} \and\\
Rowanne Trapmann\inst{1,2,3}\orcidID{0009-0000-1851-7508} \and\\
Willem-Jan van den Heuvel\inst{1,2,3}\orcidID{0000-0003-2929-413X}
}

\authorrunning{R. C. Ferreira et al.}

%%%%%%%%%%%%%%%%%%%%%%%%%%%%%%%%%%%%%%%%%%%%%%%%%%%%%%%%%%%%%%%%%%%%%%%%%%%%%%%%
%                                 AFFILIATIONS                                 %
%%%%%%%%%%%%%%%%%%%%%%%%%%%%%%%%%%%%%%%%%%%%%%%%%%%%%%%%%%%%%%%%%%%%%%%%%%%%%%%%

\institute{%
Jheronimus Academy of Data Science (JADS), 's-Hertogenbosch, The Netherlands \and
Eindhoven University of Technology (TUe), Eindhoven, The Netherlands \and
Tilburg University (TiU), Tilburg, The Netherlands
\email{r.cordeiro.ferreira@jads.nl} \\
\email{W.J.A.M.vdnHeuvel@tilburguniversity.edu} \\
\email{r.i.a.h.trapmann@tue.nl}%
}

%%%%%%%%%%%%%%%%%%%%%%%%%%%%%%%%%%%%%%%%%%%%%%%%%%%%%%%%%%%%%%%%%%%%%%%%%%%%%%%%
%                                    TITLE                                     %
%%%%%%%%%%%%%%%%%%%%%%%%%%%%%%%%%%%%%%%%%%%%%%%%%%%%%%%%%%%%%%%%%%%%%%%%%%%%%%%%

% typeset the header of the contribution
\maketitle 

%%%%%%%%%%%%%%%%%%%%%%%%%%%%%%%%%%%%%%%%%%%%%%%%%%%%%%%%%%%%%%%%%%%%%%%%%%%%%%%%
%                                  ABSTRACT                                    %
%%%%%%%%%%%%%%%%%%%%%%%%%%%%%%%%%%%%%%%%%%%%%%%%%%%%%%%%%%%%%%%%%%%%%%%%%%%%%%%%

\begin{abstract}
\vspace{-1.9em}
This case study describes challenges and lessons learned on building \OG:
a Machine Learning--Enabled System (MLES) for anomaly detection in the maritime
domain. First, the paper presents the system's specification and architecture.
\OG was designed with a microservices architecture to enable multiple teams to
work on the project in parallel. Then, the paper discusses how the developers
adapted contract-based design to MLOps for achieving that goal. As an MLES, \OG
employs code, model, and data contracts to establish guidelines between its
services. This case study hopes to inspire software engineers, machine learning
engineers, and data scientists to leverage similar approaches for their systems.


\end{abstract}

%%%%%%%%%%%%%%%%%%%%%%%%%%%%%%%%%%%%%%%%%%%%%%%%%%%%%%%%%%%%%%%%%%%%%%%%%%%%%%%%
%                                   CHAPTERS                                   %
%%%%%%%%%%%%%%%%%%%%%%%%%%%%%%%%%%%%%%%%%%%%%%%%%%%%%%%%%%%%%%%%%%%%%%%%%%%%%%%%

\section{Introduction}
\label{sec:introduction}
%%%%%%%%%%%%%%%%%%%%%%%%%%%%%%%%%%%%%%%%%%%%%%%%%%%%%%%%%%%%%%%%%%%%%%%%%%%%%%%%

% What is the deal with the maritime domain?
%------------------------------------------------------------------------------%
Maritime transportation is the cornerstone of global trade and commerce,
facilitating the movement of goods and commodities across vast distances.
The maritime domain represents one of the most dynamic and complex operational
environments in global logistics and transportation. It is characterized by
continuous movement across international boundaries, varying regulatory
frameworks, and vulnerability to unpredictable weather conditions. As such,
maritime operations generate a large amount of heterogeneous data from multiple
sources. 

% The data landscape in maritime operations is particularly diverse and
% challenging. Vessel tracking systems, such as the AIS (Automatic Identification
% System), broadcast position, speed, and navigational status every few seconds,
% generating billions of data points annually \cite{IMO2015RevisedAIS}.
% Port management systems capture cargo handling metrics, berthing schedules, and
% resource allocation information \cite{EPA2025ManagementEPA}. Regulatory bodies
% contribute compliance information, safety records, and inspection results.
% Additional data streams include crew management systems, maintenance logs, fuel
% consumption monitors, and commercial transaction records
% \cite{Tu2018ExploitingMethodology}. 

% How is data related to the maritime domain?
%------------------------------------------------------------------------------%
The digital transformation of the maritime industry has accelerated in recent
years, driven by economic pressures, environmental regulations, and safety
concerns~\cite{charamis2025theperspective}. According to the International
Maritime Organization (IMO), ships now function as floating data centers,
equipped with hundreds of sensors. Improving maritime navigation and
communication by facilitating the implementation of safe data sharing between
vessels and between vessels and shore-based systems through automation%
~\cite{Berg2015DigitalisationRe}. This complex web of information flows across
organizational boundaries between shipping companies, port authorities, maritime
safety agencies, and commercial partners~\cite{Kavallieratos2020ShippingShip}. 

% Why does the maritime domain requires machine learning / mlops?
%------------------------------------------------------------------------------%
This rich data ecosystem presents opportunities and challenges, including the
potential for predictive analytics, anomaly detection, and route optimization
using Machine Learning (ML)~\cite{Capobianco2021DeepNetworks}.
However, extracting actionable insights requires systems capable of ingesting,
processing, and analyzing different data streams at scale while maintaining
operational reliability in an environment where connectivity and conditions
rapidly change~\cite{Spiliopoulos2024PatternsPatterns}.
Machine Learning Operations (MLOps) refers to a collection of best practices,
sets of concepts, and development culture that aims to achieve that goal%
~\cite{Kreuzberger2023MachineArchitecture}.

% Why does the maritime domain requires microservices?
%------------------------------------------------------------------------------%
Since 2015, the microservices architectural style has emerged as a promising
approach to address challenges of modularity, scalability, and resilience by
decomposing complex systems into smaller, independently deployable services
with well-defined interfaces~\cite{Newman:BuildingMicroservices:2021,%
Richardson:MicroservicesPatterns:2018}.
This approach can be particularly suited to the maritime domain's nature,
where specialized components can be designed to handle different maritime data
sources and analytical needs.

% What is the relationship between microservices and mlops?
%------------------------------------------------------------------------------%
The novel field of Software Engineering for Artificial Intelligence (SE4AI)
studies the adaptation of Software Engineering techniques for the creation of
ML-Enabled Systems (MLES)~\cite{Khomh2018SoftwareAhead}. Since 2022, different
reference architectures have been proposed for them%
~\cite{Kumara2023RequirementsIndustry,Ferreira2025ASystems,%
Kreuzberger2023MachineArchitecture}. Fundamentally, they describe these systems
as containing multiple components, which can be divided across different
services.

% What is Ocean Guard?
%------------------------------------------------------------------------------%
This experience report describes the development of \OG: an \mbox{ML-Enabled}
System (MLES) for anomaly detection in the maritime domain.
  \Cref{sec:system_specification} details the functional and non-functional
  requirements considered for the system.
  \Cref{sec:system_architecture} gives an overview of the system's microservices
  architecture.
  \Cref{sec:api_architecture} focuses on the \OG API, designed with a
  hexagonal architecture using Domain-Driven Design (DDD).
  \Cref{sec:team_configuration} shows the different teams involved in
  the development of the tool.
  \Cref{sec:contract_based_development} describes how the developers adapted
  contract-based design to MLES for developing different components in parallel.
  \Cref{sec:challenges} lists the main challenges during the development
  of the \OG tool.%, according to the experience of the authors.
  \Cref{sec:lessons_learned} then summarizes the key lessons learned,
  related to applying development techniques from microservices to MLES.
  Finally, \cref{sec:conclusion} lists the main contributions of 
  the paper and future work.
  % \Cref{sec:system_infrastructure} briefly describe the main tools used
  % to provide standardized infrastructure in the project.

% Why is the paper somewhat vague?
%------------------------------------------------------------------------------%
Unfortunately, given the sensitivity of the project, this paper does not address
\OG data model or its anomaly detection techniques. Instead, it focuses on the
high-level architecture design, which may inspire other researchers and
practitioners developing similar ML-enabled systems.
\vspace{-0.3em}
\section{System Specification}
\label{sec:system_specification}
%%%%%%%%%%%%%%%%%%%%%%%%%%%%%%%%%%%%%%%%%%%%%%%%%%%%%%%%%%%%%%%%%%%%%%%%%%%%%%%%
\vspace{-0.2em}

This section describes the specification of the \OG tool. Its goal is to
analyze and detect anomalies across multiple types of data, unlike other
commercially available tools.
The section is divided into two parts:
\emph{functional} and \emph{non-functional} requirements%
~\cite{Richards:FundamentalsSoftwareArchitecture:2020}.
They were defined through interviews with the stakeholders of the project
during the inception of the \OG tool.

% Although the list of features above is not new, the project stakeholders want
% to develop \OG to create a tool capable of cross-referencing multiple types
% of data unlike other commercially available tools.

\vspace{-0.5em}
\subsection{Functional Requirements}
\label{subsec:functional_requirements}
%%%%%%%%%%%%%%%%%%%%%%%%%%%%%%%%%%%%%%%%%%%%%%%%%%%%%%%%%%%%%%%%%%%%%%%%%%%%%%%%

\Requirements{Functional Requirements} specify the observable behavior that
a system should provide~\cite{Washizaki:SWEBOK-v4:2024}.
There are two actors related to the \OG tool. The following descriptions
detail the main functionalities associated with them.

An \Actor{Investigator} is a human user who uses the tool to obtain
clues about anomalous behavior at sea. They must be able to:
%------------------------------------------------------------------------------%
\begin{enumerate}[label={I\arabic*.}, leftmargin=*]
  \item see geolocations (latitude, longitude, timestamp)
        of marine objects on a map,%
        \label{fr:i1}
  \item filter geolocations by area of interest, date and time,%
        \label{fr:i2}
  \item discern different types of marine objects
        (vessels, structures, or unidentified),%
        \label{fr:i3}
  \item retrieve geolocations gathered from different data sources,%
        \label{fr:i4}
  \item check details (metadata) associated with a given marine object,%
        \label{fr:i5}
  \item highlight the trajectory of a marine object,%
        \label{fr:i6}
  \item see anomalies identified by the tool in a map,%
        \label{fr:i7}
  \item filter anomalies by area of interest, date and time, and%
        \label{fr:i8}
  \item inspect why an anomaly was considered so by the tool.%
        \label{fr:i9}
\end{enumerate}
%------------------------------------------------------------------------------%

The \Actor{Anomaly Detector} is a component of the tool responsible for detecting
anomalous behavior of marine objects. It must be able to:
%------------------------------------------------------------------------------%
\begin{enumerate}[label={A\arabic*.}, leftmargin=*]
  \item detect anomalies related to a marine object,
        \label{fr:a1}
  \item list anomalies by area of interest, date and time, and
        \label{fr:a2}
  \item explain why an anomaly is considered so.
        \label{fr:a3}
\end{enumerate}
%------------------------------------------------------------------------------%

\subsection{Non-Functional Requirements}
\label{subsec:non-functional_requirements}
%%%%%%%%%%%%%%%%%%%%%%%%%%%%%%%%%%%%%%%%%%%%%%%%%%%%%%%%%%%%%%%%%%%%%%%%%%%%%%%%

\Requirements{Non-Functional Requirements} describe non-domain design
considerations that influence the structure of a system and are important
for its success~\cite{Richards:FundamentalsSoftwareArchitecture:2020}.
There are four main non-functional requirements proposed for the \OG tool.
The following descriptions detail the main characteristics the system
should have to attend to its users.
 
\paragraph{Explainability.}\label{nfr:explainability}
%------------------------------------------------------------------------------%
The \OG tool should be able to explain why it considers a marine object to be
anomalous. For algorithm-based anomaly detection, this means clarifying which
decisions were taken until the anomaly was identified. For AI-based anomaly
detection, this means using interpretable models (such as tree-based models) or
implementing black-box explainability techniques (such as SHAP or LIME).
  
\paragraph{Compatibility.}\label{nfr:compatibility}
%------------------------------------------------------------------------------%
The \OG tool should be able to interoperate with other existing tools, extending
the \Actor{Investigator}'s current existing capabilities. As such, the tool
should follow EU standards that enable quick integration with other ecosystems.

\paragraph{Resiliency.}\label{nfr:resiliency}
%------------------------------------------------------------------------------%
The \OG tool should be able to handle the expected amount of data.
This is important because the system may process high-volume and high-velocity
data sources, such as those providing tracking data.

\paragraph{Compliance.}\label{nfr:compliance}
%------------------------------------------------------------------------------%
The \OG tool should be compliant with the latest \mbox{European} laws, including
the General Data Protection Regulation (GDPR), and the Artificial Intelligence
Act (AI Act). To make data and model management simpler to comply with different
regulations, each stakeholder should be able to have their own installation
of \OG.
\section{System Architecture}
\label{sec:system_architecture}
%%%%%%%%%%%%%%%%%%%%%%%%%%%%%%%%%%%%%%%%%%%%%%%%%%%%%%%%%%%%%%%%%%%%%%%%%%%%%%%%

\begin{figure}[p]
  \centering
  \includegraphics[width=0.96\linewidth]{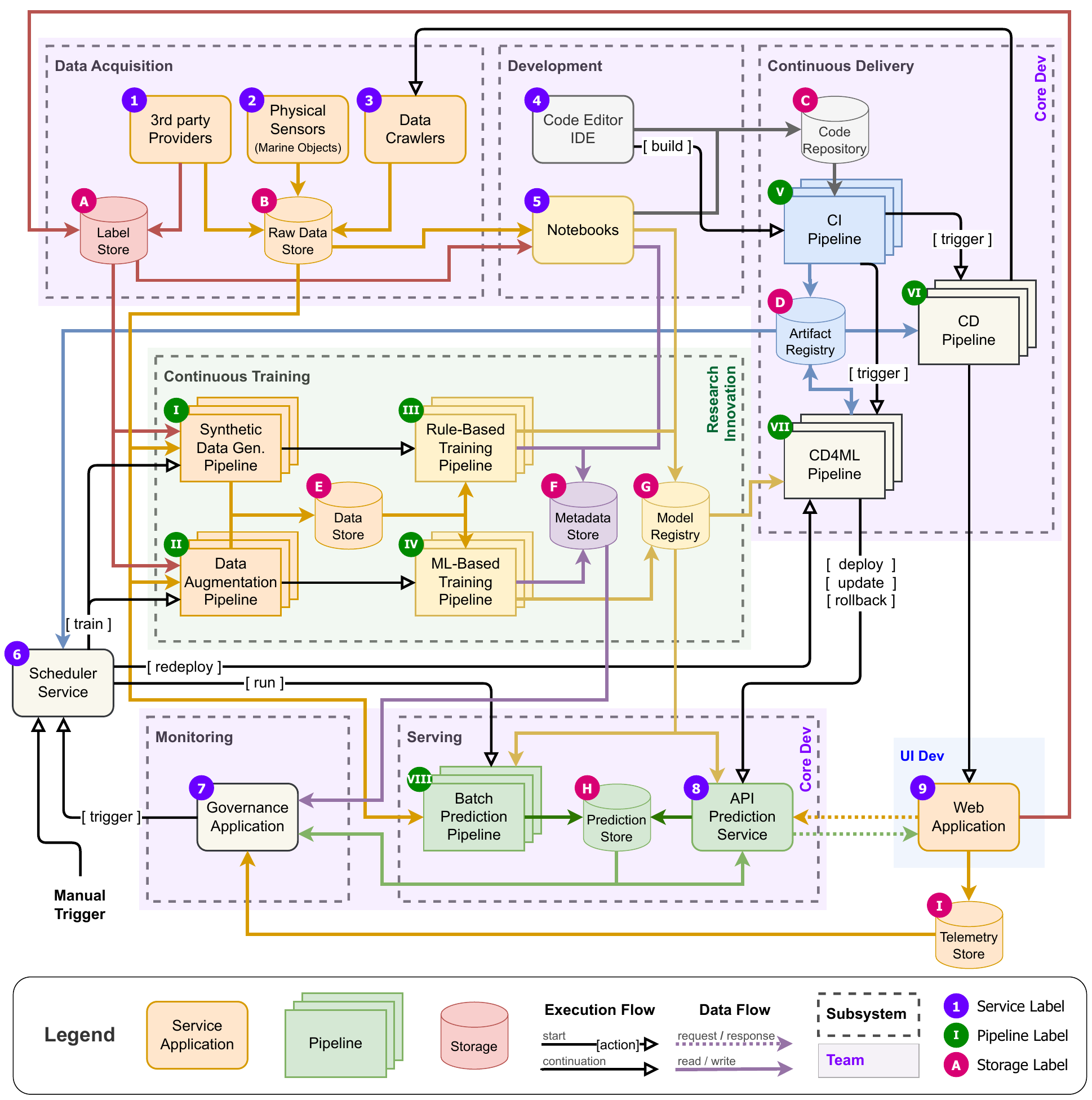}
  \caption{
    \textbf{System Architecture of the \OG tool}
    This architecture follows the same notation introduced by
    Ferreira \etal~\cite{Ferreira2025ASystems}.
    Rectangles represent either \textbf{applications} or \textbf{services},
      which execute continuously.
    Stacked rectangles represent \textbf{pipelines},
      which execute a task on demand.
    Lastly, cylinders represent \textbf{data \mbox{storage}},
      which may be databases of any type.
    Components are connected by arrows.
    Black arrows with a hollow tip illustrate the \textbf{execution flow}.
      They start and end in a component.
      Labeled arrows represent the trigger that starts a workflow,
      whereas unlabeled arrows represent the continuation of an
      existing workflow.
    Colored arrows with a filled tip illustrate the \textbf{data flow}.
    They appear in two types:
      solid arrows going to and from a data storage represent
      write and read operations, respectively;
      dotted arrows represent a sync or async request-response
      communication between components.
    Components are colored according to the data they produce:
      \mbox{\LegendColoredComponent{orange}{raw data}},
      \mbox{\LegendColoredComponent{gray}{source code}},
      \mbox{\LegendColoredComponent{blue}{executable artifacts}},
      \mbox{\LegendColoredComponent{red}{ML-specific data}},
      \mbox{\LegendColoredComponent{yellow}{ML models}},
      \mbox{\LegendColoredComponent{purple}{ML training metadata}}, and
      \mbox{\LegendColoredComponent{green}{ML model predictions}}.
      The remaining \LegendBWComponent{standalone components} orchestrate
      the execution of others.
    Components are also grouped into \textbf{subsystems},
    with their background colored according to the \textbf{teams} responsible
    for their development.
    \LegendColoredLabel{violet}{Numbers},
    \LegendColoredLabel{moss}{roman numerals} and
    \LegendColoredLabel{magenta}{letters}
    are used as labels throughout \cref{sec:system_architecture}.
  }
  \label{fig:og_system_architecture}
\end{figure}

This section presents the architecture of the \OG tool, designed to implement
the requirements presented in \cref{sec:system_specification}.
\Cref{fig:og_system_architecture} showcases the components planned for \OG,
following the reference architecture representation proposed by Ferreira
\etal~\cite{Ferreira2025ASystems}.

%------------------------------------------------------------------------------%
%------------------------------------------------------------------------------%

\paragraph{Data Acquisition.}\label{p:og_subsystem_data_acquisition}
%------------------------------------------------------------------------------%
To fulfill the requirements described in \cref{sec:system_specification},
the \OG tool required comprehensive data acquisition capabilities.
This subsystem collects data from multiple heterogeneous sources:
\LegendService{1}{3rd party Providers}, \LegendService{2}{Physical Sensors},
and \LegendService{3}{Data Crawlers}. The data is stored in one of the two
data stores: the \LegendDataStore{A}{Label Store} for known data about 
anomalies -- which may come from the \LegendService{1}{3rd Party Providers}
or the \LegendService{9}{Web Application} -- and the \LegendDataStore{B}{Raw
Data Store} for unprocessed data.

\paragraph{Continuous Training.}\label{p:og_subsystem_continuous_training}
%------------------------------------------------------------------------------%
The next step in the implementation was to enable continuous training.
Four components were developed for this purpose:
  a \LegendService{I}{Synthetic Data Generation Pipeline};
  a \LegendService{II}{Data Augmentation Pipeline};
  a \LegendService{III}{Rule-Based Training Pipeline}; and
  an \LegendService{IV}{ML-Based Training Pipeline}.
The creation of these pipelines is an opportunity to introduce two critical
model management tools:
  the \LegendDataStore{F}{Metadata Store} and
  the \LegendDataStore{G}{Model Registry},
which together facilitate model lineage tracking and versioning. 

\paragraph{Serving.}\label{p:og_subsystem_serving}
%------------------------------------------------------------------------------%
The \LegendPipeline{VIII}{Batch Prediction Pipeline} facilitates efficient
processing of large-scale offline predictions, while results are stored in a
dedicated \LegendDataStore{H}{Prediction Store}. Complementing this offline
processing capability, the \LegendService{8}{API Prediction Service} is the
main component of the system, providing low-latency inference capabilities to
support real-time application requirements. 

\paragraph{Monitoring.}\label{p:og_subsystem_monitoring}
%------------------------------------------------------------------------------%
The \OG tool requires a \LegendService{7}{Governance Application} to enable
model monitoring and analysis of system usage. This application interfaces
with a dedicated \LegendDataStore{I}{Telemetry Store} that will store data
about the use of the system while it runs.

\paragraph{Development.}\label{p:og_subsystem_development}
%------------------------------------------------------------------------------%
Currently, software engineers, data scientists, and machine learning engineers
can access the \OG source code maintained in the centralized
\LegendDataStore{C}{Code Repository}. Developers can easily modify the 
source code by employing one of the pre-configured tools:
  \LegendService{4}{Integrated Development Environments (IDEs)}
  for structured software engineering tasks,
  and \LegendService{5}{Notebooks} for interactive prototyping and analysis.
Providing standardized environments allows quick onboarding in the project
and a seamless development experience.
% This approach accommodates the diverse workflows and preferences of a
% multidisciplinary team.

\paragraph{Continuous Delivery.}\label{p:og_subsystem_continuous_delivery}
%------------------------------------------------------------------------------%
Following modern software engineering practices, a \LegendPipeline{v}{Continuous
Integration (CI) Pipeline} is used to build and test code. Thereafter, dedicated
\LegendPipeline{VI}{Continuous Delivery (CD) Pipelines} deploy traditional
software-based components, whereas specialized \LegendPipeline{VII}{Continuous
Delivery of Machine Learning (CD4ML) Pipelines} do the same for ML-based
components. These pipelines interact with a centralized
\LegendDataStore{D}{Artifact Registry} that maintains versioned software
components, containerized applications and serialized models, ensuring
reproducibility and traceability throughout the development cycle.

\section{API Architecture}
\label{sec:api_architecture}
% \vspace{-0.2em}
%%%%%%%%%%%%%%%%%%%%%%%%%%%%%%%%%%%%%%%%%%%%%%%%%%%%%%%%%%%%%%%%%%%%%%%%%%%%%%%%

This section dives deeper into the architecture of the \OG API, which
corresponds to the \LegendService{8}{API Prediction Service} in the
architecture presented in \cref{sec:system_architecture}.
The \OG API is the center of the ML-Enabled System, since it applies the
rule-based and machine-learning-based anomaly detection models over
the data received by the system.

%------------------------------------------------------------------------------%
\begin{figure}[p]
  \centering
  \includegraphics[width=1\linewidth]{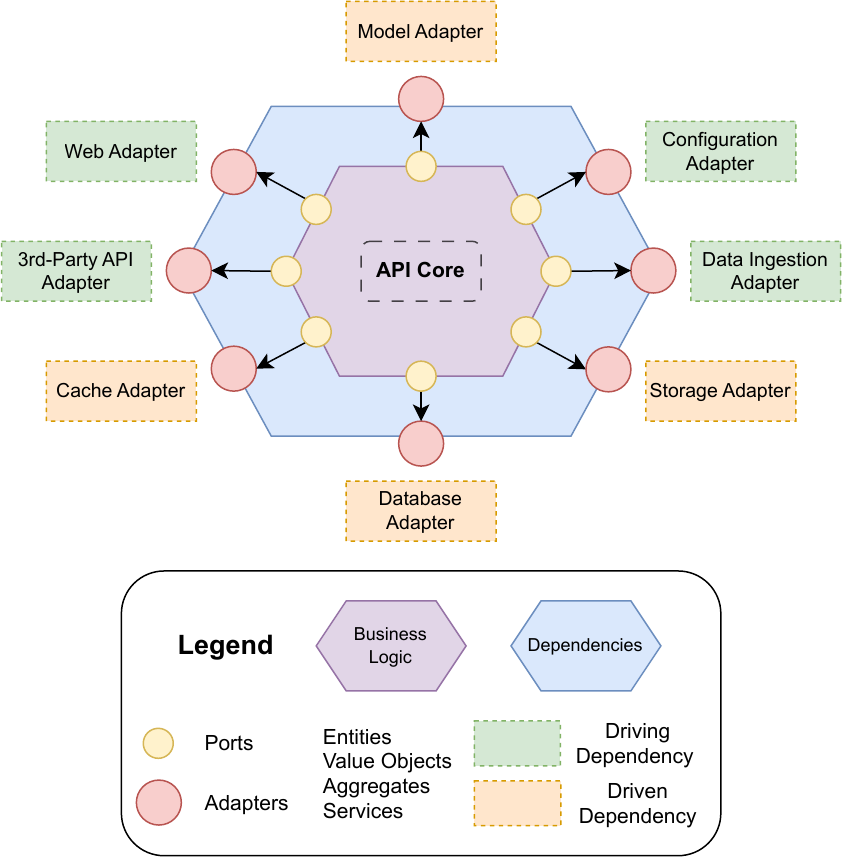}
  \caption{
    \textbf{Component Architecture of the \OG API.}
    It follows the hexagonal architecture~\cite{Martin:CleanArchitecture:2017}.
    The \LegendColoredLayer{purple}{purple hexagon}
    represents the \textbf{Business Logic},
    where Domain-Driven Design patterns such as \emph{Entities},
    \emph{Value Objects}, \emph{Aggregates}, and \emph{Services}%
    ~\cite{Evans:DomainDrivenDesign:2003} are implemented.
    The \LegendColoredLayer{blue}{blue hexagon}
    represents the \textbf{Dependencies},
    where the \emph{Ports} and \emph{Adapters} pattern is implemented
    as an \textsc{Anti-Corruption Layer} against external dependencies%
    ~\cite{Martin:CleanArchitecture:2017}.
    Rectangles illustrate the external dependencies themselves,
    with a \LegendColoredComponent{green}{green rectangle}
    representing a \textbf{Driving Dependency}
    -- which initiates actions in the API --
    while an \LegendColoredComponent{orange}{orange rectangle}
    representing a \textbf{Driven Dependency}
    -- whose actions are initiated by the API. 
    Lastly, \LegendColoredLabel{yellow}{yellow circles} represent the
    \emph{Ports} and \LegendColoredLabel{red}{red circles} represent the
    \emph{Adapters}, one per dependency. 
    Solid arrows, going from a port to an adapter represent the inside-out
    dependency flow, going from the \emph{Core} to the \emph{Ports} then
    to the \emph{Adapters}.
  }
  \label{fig:og_api_architecture}
\end{figure}
%------------------------------------------------------------------------------%

\Cref{fig:og_api_architecture} represents the key components of the API
according to a \emph{hexagonal architecture}, also known as \emph{onion} or
\emph{clean architecture}~\cite{Martin:CleanArchitecture:2017}.
This architectural pattern is designed to enhance flexibility and
maintainability of a system by ensuring that its core business logic remains
independent of external technologies.

According to the \emph{hexagonal architecture} pattern, the system can be
divided into three key elements: \emph{core}, \emph{ports}, and \emph{adapters}.
The system follows a dependency rule: As one moves inward towards the core, the
level of abstraction and the scope of policy increases. Specifically, the deeper
layers of the architecture encapsulate higher-level policies and overarching
business rules, while the outer layers handle more concrete implementation
details. The core, being the innermost layer, represents the most general and
highest level of abstraction~\cite{Martin:CleanArchitecture:2017}.

\paragraph{Core.}\label{p:og_api_core}
%------------------------------------------------------------------------------%
Responsible for implementing the business logic. The \OG API follows the
patterns of Domain-Driven Design (DDD)~\cite{Evans:DomainDrivenDesign:2003}:
\emph{entities} and \emph{value objects} represent the key objects related
to the maritime domain. The \emph{services}, also known as \emph{use cases}%
~\cite{Martin:CleanArchitecture:2017}, orchestrate the lifecycle of these
objects.

\paragraph{Ports.}\label{p:og_api_ports}
%------------------------------------------------------------------------------%
Responsible for establishing the contract and managing the communication between
the core and the adapters. Within the \OG API, key components such as database
repositories, dependency injection, security mechanisms, and web routers are
defined as \emph{ports}, facilitating the necessary integrations and ensuring
external concerns do not impact the core business logic. 

\paragraph{Adapters.}\label{p:og_api_adapters}
%------------------------------------------------------------------------------%
Responsible for communication with external dependencies, making key details
of the business logic. Adapters are categorized based on their dependencies:
reading and output. Within the \OG API, \emph{reading} adapters include
the model adapter, third-party API adapter, storage adapter, database adapter,
and configuration adapter, whereas \emph{output} adapters include the web
adapter and the cache adapter.

% \subsubsection{Database.}\label{subsubsec:og_api_database}
%------------------------------------------------------------------------------%
% The Ocean Guard API uses PostgreSQL with the PostGIS extension to support
% geospatial data management. To manage database schema changes and ensure
% version control, the Ocean Guard API uses Alembic, a flexible migration tool.

% \subsubsection{Web Server.}\label{subsubsec:og_api_webserver}
%------------------------------------------------------------------------------%
% The \OG API, implemented using the FastAPI framework, serves as the central
% component of our system architecture. This API functions as the primary
% interface between the front-end WebApp and the back-end services, handling all
% incoming requests through a structured request-response paradigm. Following
% Hexagonal Architecture principles, the API endpoints are implemented as
% adapters, effectively decoupling the core application logic from its external
% interfaces. Each endpoint is defined as a discrete Python function within the
% FastAPI routing system, which then delegates execution to the corresponding
% core service method.
\section{Team Configuration and Development Workflow}
\label{sec:team_configuration}
%%%%%%%%%%%%%%%%%%%%%%%%%%%%%%%%%%%%%%%%%%%%%%%%%%%%%%%%%%%%%%%%%%%%%%%%%%%%%%%%

As an ML-enabled system, developing the \OG tool requires multiple specialists,
including software engineers, data scientists, and machine learning engineers%
~\cite{Kreuzberger2023MachineArchitecture}. \Cref{tab:og_teams} summarizes
the project  organizational structure, which consists of four teams working in
parallel, yet collaborative streams.

%------------------------------------------------------------------------------%
\begin{table}[t!]
  \centering
  \includegraphics[width=\linewidth]{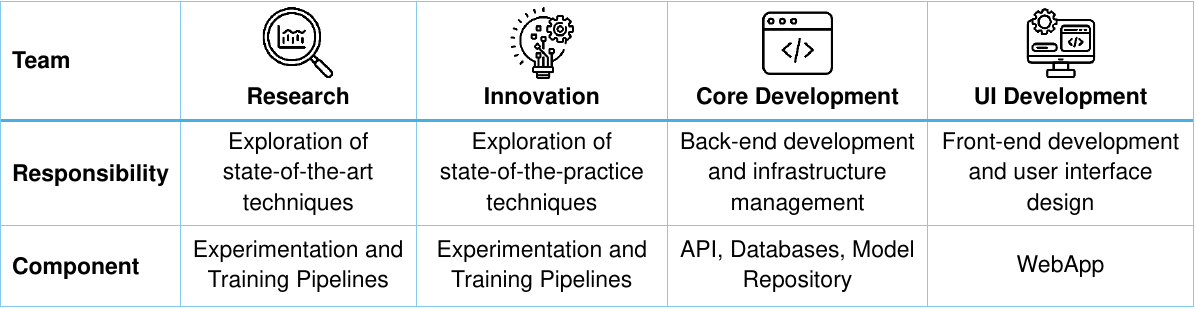}
  \vspace{-0.3em}%
  \caption{
    \textbf{\OG Teams.}
    The \Team{Research} and \Team{Innovation} teams -- composed of M.Sc.
    students and Eng.D. trainees -- focus on different AI-based and
    algorithmic techniques. Whilst the work from them is research- and
    innovation-oriented, their work directly contributes to the \OG API Model
    Adapter. The \Team{Core Development} team focuses on the \Component{API}
    whereas the \mbox{\Team{UI Development}} focuses on the \Component{WebApp}. 
  }
  \label{tab:og_teams}
  \vspace{-0.6em}
\end{table}
%------------------------------------------------------------------------------%

The project workflow is deliberately structured to minimize interdependency
between teams. Each team operates within clearly defined boundaries whose goal
is to allow parallel development without blocking dependencies. The integration
points happen via the CI/CD pipeline~\cite{Humble:ContinuousDelivery:2010},
where independently developed components can be tested against predefined
contracts before merging into the main development branch%
~\cite{Richardson:MicroservicesPatterns:2018}.
This design technique is further explored in
\cref{sec:contract_based_development}.

% \vspace{-0.1em}
\paragraph{Research Team.}\label{team:research}
%------------------------------------------------------------------------------%
It consists of Master of Science (M.Sc.) students completing their theses.
This team primarily focuses on exploring state-of-the-art techniques in the
maritime domain. Although research-oriented, their work directly contributes
data products to the \Component{API}, particularly to the API Model Adapter. 

% \vspace{-0.1em}
\paragraph{Innovation Team.}\label{team:innovation}
%------------------------------------------------------------------------------%
It consists of Engineering Doctorate (Eng.D.) trainees focusing on the
exploration and implementation of state-of-the-practice techniques. This team
employs rapid prototyping methodologies within constrained timelines to deliver
demonstrable improvements to the API Model Adapter.

% \vspace{-0.1em}
\paragraph{Core Development Team.}\label{team:core_development}
%------------------------------------------------------------------------------%
It consists of the authors of this paper focusing on the fundamental aspects of
the project infrastructure, including back-end development and infrastructure
management. This team focuses on the \Component{API}, databases, and model
repository, with particular emphasis on refining the central \Component{API}
that serves as the integration point for all microservices. 
 
% \vspace{-0.1em}
\paragraph{UI Development.}\label{team:ui_development}
%------------------------------------------------------------------------------%
It consists of hired web developers specializing in user interface design and
front-end development. This team focuses primarily on the \Component{WebApp},
which will be used by the end users for maritime tracking activities.

% \vspace{-0.1em}
\bigskip
%------------------------------------------------------------------------------%
As the \Team{Research} and \Team{Innovation} teams consist of students and
trainees, they operate under specific academic deadlines. These external
constraints have been factored into the development process, further reinforcing
the decision to implement a parallel work approach. 

\section{Contract-Based Development}
\label{sec:contract_based_development}
%%%%%%%%%%%%%%%%%%%%%%%%%%%%%%%%%%%%%%%%%%%%%%%%%%%%%%%%%%%%%%%%%%%%%%%%%%%%%%%%

% intro
% code, data, model contracts
% Research and Innovation -> Data Schema 
% Front-end -> API, endpoints 

\OG is a complex system, as discussed in~\cref{sec:system_architecture}.
\emph{Design by Contract} is a recommended modeling technique for systems
with such an architecture~\cite{Ford:SoftwareArchitectureHardParts:2021}.
It is an approach that establishes clear interfaces and agreements between
software components or teams, defining how they interact without requiring
knowledge of each other's internal implementation%
~\cite{Ford:SoftwareArchitectureHardParts:2021}.

In the development of the \OG tool, contract-based development was
systematically implemented across the boundaries of interaction of the four
teams introduced at \cref{sec:team_configuration}. These contracts serve as
a formal specification without constraining the internal implementations%
~\cite{Richardson:MicroservicesPatterns:2018}. As the project evolves,
contracts can be refined to keep teams aligned.

\subsection{Code Contracts}
\label{subsec:code_contracts}
%%%%%%%%%%%%%%%%%%%%%%%%%%%%%%%%%%%%%%%%%%%%%%%%%%%%%%%%%%%%%%%%%%%%%%%%%%%%%%%%

For this project, \emph{code contracts} document the expected behavior between
two services -- owned by different teams -- that interact synchronously or
asynchronously via the network, usually under the HTTP protocol. As such, these
specifications include endpoints, request parameters, and expected responses.

There are two main \emph{code contracts} in the \OG tool:
%------------------------------------------------------------------------------%
\begin{itemize}
  \setlength{\itemsep}{\baselineskip}
  \item Between \LegendService{3}{Data Crawlers} and external data sources.
        Creating these contracts provides the \Team{Core Development} team
        with an anti-corruption layer~\cite{Evans:DomainDrivenDesign:2003}
        against the external data sources. This allows them to create tests
        that verify the expected behavior of 3rd-party APIs, which may change
        without notice.
  \item Between the \LegendService{8}{API Prediction Service} and the 
        \LegendService{8}{Web Application}.
        Creating this contract allows the \Team{UI Development} team
        to mock the responses of the \LegendService{8}{API Prediction Service}.
        This allows them to develop the front end in parallel with the
        back end developed by the \Team{Core Development} team.
\end{itemize}
%------------------------------------------------------------------------------%

% We formalized contracts as request-response specifications for the API
% interactions. These specifications detail the endpoints, required parameters,
% and expected response structure. This approach facilitated the simultaneous
% development of the WebApp component and the Core API, eliminating dependencies
% that might otherwise have created bottlenecks in the development process. 

\subsection{Data Contracts}
\label{subsec:data_contracts}
%%%%%%%%%%%%%%%%%%%%%%%%%%%%%%%%%%%%%%%%%%%%%%%%%%%%%%%%%%%%%%%%%%%%%%%%%%%%%%%%

For this project, \emph{data contracts} document the expected formats in a
data storage whose data is produced and consumed by services or pipelines
owned by different teams. As such, these specifications include types, formats,
distributions (when applicable or known) of data, besides the protocol for
reading and writing the data itself.

\break
There are three main \emph{data contracts} in the \OG tool:%
%------------------------------------------------------------------------------%
\begin{itemize}
  \setlength{\itemsep}{0.5\baselineskip}
  \item At the \LegendDataStore{A}{Label Store},
        between the
          \LegendService{1}{3rd Party Applications} and 
          \LegendService{9}{Web Application}
        that produce them, and the
          \LegendService{5}{Notebooks},
          \LegendPipeline{I}{Synthetic Data Generation Pipeline} and
          \LegendPipeline{II}{Data Augmentation Pipeline}
        that consume them.
        These contracts ensure the \Team{Research} and \Team{Innovation} teams
        know what to expect to process labels.

  \item At the \LegendDataStore{B}{Raw Data Store},
        between the
          \LegendService{1}{3rd Party Applications},
          \LegendService{2}{Physical Sensors} and
          \LegendService{3}{Data Crawlers}
        that produce them, and the
          \LegendService{5}{Notebooks},
          \LegendPipeline{I}{Synthetic Data Generation Pipeline},
          \LegendPipeline{II}{Data Augmentation Pipeline} and
          \LegendPipeline{VIII}{Batch Prediction Pipeline}
        that consume them.
        These contracts ensure the \Team{Research} and \Team{Innovation} teams
        know what to expect to process raw data, as well as what their models
        might use during prediction.

  \item At the \LegendDataStore{E}{Data Store},
        between the
          \LegendPipeline{I}{Synthetic Data Generation Pipeline} and
          \LegendPipeline{II}{Data Augmentation Pipeline}
        that produce them, and the
          \LegendPipeline{III}{Rule-Based Training Pipeline} and
          \LegendPipeline{IV}{ML-Based Training Pipeline} pipelines
        that consume them.
        These contracts ensure the \Team{Research} and \Team{Innovation} teams
        know how to consume the post-processed raw data and labels created
        by each other.
\end{itemize}
%------------------------------------------------------------------------------%

% For the components developed by the \Team{Research} and \Team{Innovation}
% teams, in particular \LegendPipeline{I}{Synthetic Data Generation} and
% \LegendPipeline{II}{Data Augmentation} pipelines, the key point we established
% data schema that served as contractual boundaries. 

\subsection{Model Contracts}
\label{subsec:model_contracts}
%%%%%%%%%%%%%%%%%%%%%%%%%%%%%%%%%%%%%%%%%%%%%%%%%%%%%%%%%%%%%%%%%%%%%%%%%%%%%%%%

For this project, \emph{model contracts} document the expected input and output
of a rule-based or ML-based model, as well as the format in which a trained
model should be stored. As such, these specifications include data types,
file formats, and data scientist--defined hyperparameters.

There is one main \emph{model contract} in the \OG tool,
%------------------------------------------------------------------------------%
at the \LegendDataStore{G}{Model Registry},
between the
  \LegendPipeline{III}{Rule-Based Training Pipeline} and
  \LegendPipeline{IV}{ML-Based Training Pipeline} pipelines
that create them, and both the
  \LegendPipeline{VIII}{Batch Prediction Pipeline} and
  \LegendService{9}{API Prediction Service}
that use them.
%------------------------------------------------------------------------------%

These contracts ensure the \Team{Research} and \Team{Innovation} teams
know how to design their models for the \Team{Core Development} team to
use in the \OG tool back-end.
This way, all teams know what the model may expect as input, what it should
return as output, and which artifacts it may produce while executing. These
contracts ensure seamless integration of the model during runtime.

% Endpoints and the HTTP request/response contract HTTP-based contracts.
% Each contract consists of an HTTP request and an HTTP reply.
% These are the endpoints.

\section{Challenges}
\label{sec:challenges}
%%%%%%%%%%%%%%%%%%%%%%%%%%%%%%%%%%%%%%%%%%%%%%%%%%%%%%%%%%%%%%%%%%%%%%%%%%%%%%%%

In the experience of the authors, there were three main challenges for the
development of the \OG tool: \emph{coupling}, \emph{alignment}, and
\emph{communication}.

\paragraph{Coupling.}\label{cl:coupling} % entanglement -> data, data model
%------------------------------------------------------------------------------%
The \OG tool was intrinsically complex, as discussed in
\cref{sec:system_architecture}. Any modification to one of its components --
either on the code, model, or data dimension -- can quickly cascade through
the system. Therefore, mitigating the drawbacks of coupling was an important
architectural challenge, particularly since services were being implemented
concurrently during development.

% As a consequence, the authors designed a microservices architectural
% style, using (code, data, model) contracts as described in
% \cref{sec:contract_based_development} to mitigate coupling between each service.

\paragraph{Alignment.}\label{cl:alignment}
%------------------------------------------------------------------------------%
The \OG tool required specialized developers, as
discussed in \cref{sec:team_configuration}. Four teams worked simultaneously on
different components, integrated via the \OG API. Therefore, keeping all team
members oriented towards the same goal was an important organizational
challenge, particularly due to the time-consuming nature of coordination
activities.

\paragraph{Communication.}\label{cl:communication}
%------------------------------------------------------------------------------%
The \OG tool had a diverse set of requirements, as discussed in
\cref{sec:system_specification}. Therefore, communicating the evolution
of the \OG tool to internal and external stakeholders -- with diverse
technical backgrounds -- was an important project management challenge, 
particularly since there were frequent feedback sessions due to its agile
development process.

% \paragraph{Coordination.}\label{cl:coordination}
% %------------------------------------------------------------------------------%
% As an ML-enabled system, the \OG tool required specialized developers, as
% discussed in \cref{sec:team_configuration}. Therefore, the team grew in size.
% This expansion made orchestrating the team much more critical:
% from its inception, the authors employed (software, data, model) contracts
% to facilitate this coordination, as discussed in
% \cref{sec:contract_based_development}.
\vspace{-0.1em}
\section{Lessons Learned}
\label{sec:lessons_learned}
%%%%%%%%%%%%%%%%%%%%%%%%%%%%%%%%%%%%%%%%%%%%%%%%%%%%%%%%%%%%%%%%%%%%%%%%%%%%%%%%

The challenges described in \cref{sec:challenges} led to two lessons learned
related to applying techniques from the microservices literature that can be
suitable for ML-enabled system development: \emph{contract-based design} and
\emph{ubiquitous language}.

\paragraph{Contract-Based Design.}\label{ll:contract-based_design}
%------------------------------------------------------------------------------%
As previously mentioned in \cref{sec:contract_based_development},
this project employed \emph{design by contract}%
~\cite{Ford:SoftwareArchitectureHardParts:2021} during the
development of the \OG tool, a recommended modeling technique for systems
with a microservices architecture~\cite{Richardson:MicroservicesPatterns:2018}.
This approach facilitated the definition of clear and precise specifications
between components, setting expectations for interfaces and behaviors.

By applying design by contract through the development process, it tackled two
challenges above: \emph{coupling}, by reducing potential integration issues and
improving system cohesion and \emph{alignment}, making it easier for the team
to discuss how to integrate the components of the \OG tool.

Applying \emph{design by contract} for MLOps is a contribution of this paper.
To develop \OG, this project took the idea of applying code contracts between
microservices~\cite{Richardson:MicroservicesPatterns:2018} to also
use data and model contracts between components of MLES.

\paragraph{Ubiquitous Language.}\label{ll:ubiquitous_language}
%------------------------------------------------------------------------------%
As previously mentioned in \cref{sec:system_architecture}, this project employed
\emph{Domain-Driven Design (DDD)}~\cite{Evans:DomainDrivenDesign:2003} during
the development of the \OG tool, a recommended modeling technique for systems
with a microservices architecture~\cite{Newman:BuildingMicroservices:2021}.
This approach facilitated the creation of a shared vocabulary between developers
and stakeholders, grounding discussions about the evolution of the system.

By refining a ubiquitous language through the development process, it tackled
two challenges above: \emph{communication}, by deepening stakeholders' and
developers' understanding of the system and improving delivery feedback and
\emph{alignment}, making it easier for the team to propose how to evolve the
components of the \OG tool.

Creating a \emph{ubiquitous language} for MLOps is a contribution of this paper.
To develop \OG, this project took the idea of sharing common terms between bounded
contexts~\cite{Evans:DomainDrivenDesign:2003} to also explain data- and
model-related concepts for the stakeholders.

\section{Conclusions}
\label{sec:conclusion}
%%%%%%%%%%%%%%%%%%%%%%%%%%%%%%%%%%%%%%%%%%%%%%%%%%%%%%%%%%%%%%%%%%%%%%%%%%%%%%%%

As the digital transformation of the maritime industry has accelerated, ships
now function as floating data centers equipped with hundreds of sensors.

To extract actionable insights from this vast amount of data, this paper
describes the \OG tool: a Machine Learning-Enabled System (MLES) for anomaly
detection in the maritime domain. Its microservices architecture enables
multiple specialized teams to work on the project in parallel, facilitated by a
contract-based design approach.
 
Developing the \OG tool presented three major challenges:
%------------------------------------------------------------------------------%
\begin{itemize}
  \setlength{\itemsep}{0.5\baselineskip}
  \item \emph{coupling},
        related to cascading changes throughout the system;
  \item \emph{alignment},
        related to coordinating four specialized teams working concurrently; and
  \item \emph{communication},
        related to explaining the evolving system to stakeholders with diverse
        technical backgrounds.
\end{itemize}
%------------------------------------------------------------------------------%
To address them, the authors applied two techniques from the microservices
literature during development:
%------------------------------------------------------------------------------%
\begin{itemize}
  \setlength{\itemsep}{0.5\baselineskip}
  \item \emph{contract-based design},
        which established clear specifications between components and reduced
        integration issues; and
  \item \emph{ubiquitous language},
        which created a shared vocabulary among all participants via the
        application of Domain-Driven Design (DDD).
\end{itemize}
%------------------------------------------------------------------------------%

\emph{Contract-based design} effectively mitigated \emph{coupling} between
components and improved \emph{alignment} among the team, whereas the
\emph{Ubiquitous Language} enhanced \emph{communication} with stakeholders and
further strengthened \emph{alignment} among the team.
Therefore, this case study in the maritime domain illustrated how these
software engineering techniques can be adapted for MLOps by extending
their applicability to the data and model dimensions of complex
Machine Learning--Enabled Systems.
% Both techniques were adapted for MLOps by extending their applicability on
% microservices to the data and model dimension of Machine Learning--Enabled Systems.
% Therefore, this case study in the maritime domain illustrates how established
% software engineering practices can be successfully adapted to complex Machine
% Learning--Enabled Systems. 

Future work will focus on continuing the development of the \OG tool to ensure
all functional and non-functional requirements are met. The development process
will remain guided by the established contract-based approach and shared
ubiquitous language. Moreover, the developers will continue to explore and
implement both state-of-the-art and state-of-the-practice techniques with the
\Team{Research} and \Team{Innovation} teams. These parallel exploration streams
will enable the \OG tool to evolve with the latest advancements in maritime
anomaly detection.
% Specific areas for future development include enhancing the model's
% adaptability to new sensor configurations, improving anomaly detection through
% advanced algorithms, and optimizing the system architecture.
% \input{chapters/X-system_infrastructure}

%%%%%%%%%%%%%%%%%%%%%%%%%%%%%%%%%%%%%%%%%%%%%%%%%%%%%%%%%%%%%%%%%%%%%%%%%%%%%%%%
%                                 BIBLIOGRAPHY                                 %
%%%%%%%%%%%%%%%%%%%%%%%%%%%%%%%%%%%%%%%%%%%%%%%%%%%%%%%%%%%%%%%%%%%%%%%%%%%%%%%%

\clearpage
\bibliographystyle{splncs04}
\bibliography{mendeley,references_maritime}

\end{document}